\documentclass[10pt]{article}
\usepackage{amssymb,amsmath}
\usepackage{amsthm}
\usepackage{mathrsfs}  
\usepackage{graphicx}
\usepackage{dcolumn}
\usepackage{bm}
\usepackage{fullpage}
\usepackage{color}
\usepackage[all]{xy}
\begin{document}
\newtheorem{Def}{Definition}[section]
\newtheorem{Thm}{Theorem}[section]
\newtheorem{Proposition}{Proposition}[section]
\newtheorem{Lemma}{Lemma}[section]
\theoremstyle{definition}
\newtheorem*{Proof}{Proof}
\newtheorem{Example}{Example}[section] 
\newtheorem{Postulate}{Postulate}[section]
\newtheorem{Corollary}{Corollary}[section]
\newtheorem{Remark}{Remark}[section]
\theoremstyle{remark}
\newcommand{\beq}{\begin{equation}}
\newcommand{\beqa}{\begin{eqnarray}}
\newcommand{\eeq}{\end{equation}}
\newcommand{\eeqa}{\end{eqnarray}}
\newcommand{\non}{\nonumber}
\newcommand{\fr}[1]{(\ref{#1})}
\newcommand{\cc}{\mbox{c.c.}}
\newcommand{\nr}{\mbox{n.r.}}
\newcommand{\bb}{\mbox{\boldmath {$b$}}}
\newcommand{\bbe}{\mbox{\boldmath {$e$}}}
\newcommand{\bt}{\mbox{\boldmath {$t$}}}
\newcommand{\bn}{\mbox{\boldmath {$n$}}}
\newcommand{\br}{\mbox{\boldmath {$r$}}}
\newcommand{\bC}{\mbox{\boldmath {$C$}}}
\newcommand{\bH}{\mbox{\boldmath {$H$}}}
\newcommand{\bp}{\mbox{\boldmath {$p$}}}
\newcommand{\bx}{\mbox{\boldmath {$x$}}}
\newcommand{\bF}{\mbox{\boldmath {$F$}}}
\newcommand{\bT}{\mbox{\boldmath {$T$}}}
\newcommand{\bomega}{\mbox{\boldmath {$\omega$}}}
\newcommand{\ve}{{\varepsilon}}
\newcommand{\e}{\mathrm{e}}
\newcommand{\F}{\mathrm{F}}
\newcommand{\Loc}{\mathrm{Loc}}
\newcommand{\Ree}{\mathrm{Re}}
\newcommand{\Imm}{\mathrm{Im}}
\newcommand{\hF}{\widehat F}
\newcommand{\hL}{\widehat L}
\newcommand{\tA}{\widetilde A}
\newcommand{\tB}{\widetilde B}
\newcommand{\tC}{\widetilde C}
\newcommand{\tL}{\widetilde L}
\newcommand{\tK}{\widetilde K}
\newcommand{\tX}{\widetilde X}
\newcommand{\tY}{\widetilde Y}
\newcommand{\tU}{\widetilde U}
\newcommand{\tZ}{\widetilde Z}
\newcommand{\talpha}{\widetilde \alpha}
\newcommand{\te}{\widetilde e}
\newcommand{\tv}{\widetilde v}
\newcommand{\ts}{\widetilde s}
\newcommand{\tx}{\widetilde x}
\newcommand{\ty}{\widetilde y}
\newcommand{\ud}{\underline{\delta}}
\newcommand{\uD}{\underline{\Delta}}
\newcommand{\chN}{\check{N}}
\newcommand{\cA}{{\cal A}}
\newcommand{\cB}{{\cal B}}
\newcommand{\cC}{{\cal C}}
\newcommand{\cD}{{\cal D}}
\newcommand{\cF}{{\cal F}}
\newcommand{\cI}{{\cal I}}
\newcommand{\cL}{{\cal L}}
\newcommand{\cM}{{\cal M}}
\newcommand{\cR}{{\cal R}}
\newcommand{\cS}{{\cal S}}
\newcommand{\cY}{{\cal Y}}
\newcommand{\cZ}{{\cal Z}}
\newcommand{\cU}{{\cal U}}
\newcommand{\cV}{{\cal V}}
\newcommand{\tcA}{\widetilde{\cal A}}
\newcommand{\DD}{{\cal D}}
\newcommand\TYPE[3]{ \underset {(#1)}{\overset{{#3}}{#2}}  }
\newcommand{\bfe}{\boldsymbol e} 
\newcommand{\bfb}{{\boldsymbol b}}
\newcommand{\bfd}{{\boldsymbol d}}
\newcommand{\bfh}{{\boldsymbol h}}
\newcommand{\bfj}{{\boldsymbol j}}
\newcommand{\bfn}{{\boldsymbol n}}
\newcommand{\bfA}{{\boldsymbol A}}
\newcommand{\bfB}{{\boldsymbol B}}
\newcommand{\bfJ}{{\boldsymbol J}}
\newcommand{\dr}{\mathrm{d}}
\newcommand{\TE}{\mathrm{TE}}
\newcommand{\TM}{\mathrm{TM}}
\newcommand{\Ai}{\mathrm{Ai}}
\newcommand{\Bi}{\mathrm{Bi}}
\newcommand{\sech}{\mathrm{sech}}
\newcommand{\jthree}{  \TYPE 3  {j}  {}   }
\newcommand{\Lam}{ \TYPE q  {\Lambda}   {}   }
\newcommand{\alp}{ \TYPE p  {\alpha}   {}   }
\newcommand{\al}[1]{ \TYPE {#1}  {\alpha}   {}   }
\newcommand{\bep}{ \TYPE p  {\beta}   {}   }
\newcommand{\be}[1]{ \TYPE {#1}  {\beta}   {}   }
\newcommand{\gamq}{ \TYPE q  {\gamma}   {}   }
\newcommand{\hash}{\#}
\newcommand{\hashat}{\widehat{\#}}
\newcommand{\hashch}{\stackrel{\vee}{\#}}
\newcommand{\chd}{\stackrel{\vee}{\D}}
\newcommand\NN[1]{{\cal N}_{#1}}
\newcommand\MM[1]{{\cal M}_{#1}}
\newcommand\BAE[1]{{\begin{equation}{\begin{aligned}#1\end{aligned}}\end{equation}}}
\newcommand{\GamCLamM}[1]{{\Gamma\mathbb{C}\Lambda^{{#1}}\cal{M}}}
\newcommand{\GamLamM}[1]{{\Gamma\Lambda^{{#1}}\,\cal{M}}}
\newcommand{\GamLamU}[1]{{\Gamma\Lambda^{{#1}}\,\cal{U}}}
\newcommand{\GamLamHU}[1]{{\Gamma\Lambda^{{#1}}\,\widehat{\cal{U}}}}
\newcommand{\GamLam}[2]{{\Gamma\Lambda^{{#1}}\,{#2}}}
\newcommand{\GTM}{{\Gamma T\cal{M}}}
\newcommand{\GTU}{{\Gamma T\cal{U}}}
\newcommand{\GT}[1]{{\Gamma T {#1}}}
\newcommand{\normM}[2]{\left(  #1\, , \, #2 \right)}
\newcommand{\normU}[2]{\left\{ #1\, , \, #2 \right\}}
\newcommand{\diag}[1]{\mbox{diag}\{\, #1\,\}}
\newcommand{\GtM}[2]{\Gamma T^{#1}_{#2}{\cal M}}
\newcommand{\inp}[2]{\left\langle\,  #1\, , \, #2\, \right\rangle}
\newcommand{\defi}{\noindent {\bf Definition : } }
\newcommand{\prop}{\noindent {\bf Proposition : } }
\newcommand{\theo}{\noindent {\bf Theorem : } }
\newcommand{\exam}{\noindent {\bf Example : } }
\newcommand{\equp}[1]{\overset{\mathrm{#1}}{=}}
\newcommand{\wt}[1]{\widetilde{#1}}
\newcommand{\wh}[1]{\widehat{#1}}
\newcommand{\ch}[1]{\check{#1}}
\newcommand{\ii}{\imath}
\newcommand{\ic}{\iota}
\newcommand{\mi}{\,\mathrm{i}\,}
\newcommand{\mr}{\,\mathrm{r}\,}
\newcommand{\mbbC}{\mathbb{C}}
\newcommand{\mbbR}{\mathbb{R}}
\newcommand{\mbbZ}{\mathbb{Z}}
\newcommand{\Leftrightup}[1]{\overset{\mathrm{#1}}{\Longleftrightarrow}}
\newcommand{\ol}[1]{\overline{#1}}
\newcommand{\rmC}{\mathrm{C}}
\newcommand{\rmH}{\mathrm{H}}
\newcommand{\Id}{\mathrm{Id}} 
\title{
Maps on statistical manifolds 
exactly reduced from the Perron-Frobenius equations  
for solvable chaotic maps 
}
\author{  Shin-itiro GOTO and  Ken UMENO,\\ 
Department of Applied Mathematics and Physics, 
Graduate School of Informatics, \\
Kyoto University, Yoshida Honmachi, Sakyo-ku, Kyoto, 606-8501, Japan
} 
\date{\today}
\maketitle
\begin{abstract}%
Maps on a parameter space for 
expressing distribution functions are  
exactly derived from the Perron-Frobenius equations for 
a generalized Boole transform family.  
Here the generalized Boole transform family 
is a one-parameter family of maps where it is defined 
on a subset of the real line and its probability distribution function  
is the Cauchy distribution with some parameters.  
With this reduction, some relations between 
the statistical picture and the orbital one are shown. 
From the viewpoint of information geometry, 
the parameter space can be identified with   
a statistical manifold, and then  
it is shown that the derived  maps can be characterized. 
Also, with an induced symplectic structure from a statistical structure, 
symplectic and information geometric 
aspects of the derived maps are discussed.   
\end{abstract}%
\section{Introduction}
Solvable chaotic maps are maps whose invariant measures
are analytically expressed, and these maps play various roles 
in physics,  applied mathematics and its engineering applications, since 
they provide analytic formulae for correlation functions and 
some average quantities \cite{Umeno1997}. With the solvable features 
one can investigate mathematical properties analytically.    
Aside from its purely academic interest, its resolution has some 
implications \cite{Umeno2000}. 
In Ref.\,\cite{Umeno2016}, a one-parameter family of maps called  
the generalized Boole transform family 
was proposed, and  
the Lyapunov exponent for 
this family was analytically obtained \cite{Umeno-Okubo2016}.   
In Ref.\,\cite{Shintani-Umeno2017} the generalized Boole transform with 
a particular parameter was used as a toy model for clarifying mechanism 
of a class of synchronizations. 
Since the generalized Boole transform family has a mixing property,  
the long-time limit of distribution functions can be estimated. 
Thus, the next interest for us is 
to describe relaxation processes of these maps by analyzing the 
Perron-Frobenius equations. Here the Perron-Frobenius equation describes 
dynamics of distribution functions, and it is not necessary to give   
a finite dimensional description of dynamics of distribution functions 
even for 
solvable chaotic systems.  If the Perron-Frobenius equation reduces to 
a finite dimensional map, 
then one can easily handle some relaxation processes.  
Thus, it is expected that such a reduction gives us various benefits. 

Information geometry is a geometrization of mathematical statistics \cite{AN}, 
and various mathematical statements have been found. This geometry gives 
tools to study statistical quantities defined on statistical manifolds, 
where statistical manifolds are identified as parameter spaces 
for parametric distribution 
functions. 
Examples of applications of information geometry include 
statistical interference, 
quantum information, 
and thermodynamics 
\cite{AN,Ay,Hayashi}.  
From these examples, one sees that 
the application of information geometry to sciences and engineering 
enables one 
to visualize theories and 
to utilize differential geometric tools for their analysis.
Thus it is expected that enlarging the application area of 
information geometry can brings various benefits. 
Moreover, if the dimension of a statistical manifold is even, then 
symplectic geometry is of interest since 
symplectic geometry provides a set of comprehensive tools to 
understand dynamical systems. As an example, Darboux's theorem 
guarantees the existence of canonical coordinates. Thus,  
a compatibility of symplectic geometry and information geometry 
is of interest \cite{LZ2017}. To state such a compatibility,      
the conditions that a statistical manifold 
admits a symplectic structure have been studied \cite{Noda2011}. 
Also it should be noted that dynamical systems on statistical manifolds 
have been studied in the literature 
\cite{Fujiwara-Amari1995, Nakamura1994,Uwano2016, Noda2016}. 

In this paper, dynamics of distribution functions 
for the generalized Boole transform family    
is focused with the Perron-Frobenius equations,  
and a family of maps is derived, where this family is defined 
on a parameter space for parametric  
distribution functions associated with the transform family.   
Then it is shown that these derived maps are characterized with tools in 
information and symplectic geometries.   
In this way, the derived family of maps is geometrically formulated.

\section{Generalized Boole transform}
In this section, the generalized Boole transform is introduced and 
some basic properties are summarized. 

The following one-parameter family of maps is focused in this paper.
\begin{Def}
(The Generalized Boole transform, \cite{Aaronson1997}).   
Let $\cR_{\,-}=\mbbR\setminus\cR_{\,-}^{\,\infty}$ be a subset of 
the real line $\mbbR$ with $\cR_{\,-}^{\,\infty}$ specified later, 
$\alpha>0$ a real number, 
and $F_{\,\alpha}: \cR_{\,-}\to\cR_{\,-}$ the map such that 
\beq
F_{\,\alpha}(\xi)
=\alpha\left(\xi-\frac{1}{\xi}\right).
\label{definition-generalized-Boole-transform-alpha}
\eeq
The set $\cR_{\,-}^{\,\infty}$ is a collection of points so that $F_{\,\alpha}(\xi)$ 
is finite for all points of $\cR_{\,-}$. Then the  map with $\alpha$,    
$F_{\,\alpha}:\cR_{\,-}\to\cR_{\,-}$ 
is referred to as the generalized Boole transform. 
Also, the one-parameter family of maps $\{F_{\,\alpha}\}$ is  referred to as the 
generalized Boole transform family.
\end{Def}
When treating $F_{\,\alpha}$ 
as a map of a dynamical system, we write it as $\xi\mapsto F_{\,\alpha}(\xi)$ 
or $\xi_{\,n+1}=F_{\,\alpha}(\xi_{\,n})$ where $\xi_{\,n}\in\cR_{\,-}$ with $n\in\mbbZ$.

One can generalize this family of maps further. 
One possible generalization is 
$F_{\,\alpha,\beta}(\xi)=\alpha\, \xi-\beta/\xi$ with some $\beta\in\mbbR$. 
In this case after introducing changes of variables 
one can show that $F_{\,\alpha,\beta}$ reduces to $F_{\,\alpha}$. 
Another generalization is found in Ref.\,\cite{Umeno2016}.

With $\{F_{\,\alpha}\}$, one can have a family of one-dimensional 
dynamical systems on $\cR_{\,-}$. This is stated as follows.  
\begin{Proposition}
(Invariant measure of the generalized Boole transform, 
\cite{Umeno-Okubo2016}).   
The dynamical system $(\cR_{\,-},\mu_{\,\alpha},F_{\,\alpha})$ with $0<\alpha<1$ 
has a mixing property with the invariant measure 
$$
\mu_{\,\alpha}(\dr \xi)
=C\left(\,\xi;0,\sqrt{\frac{\alpha}{1-\alpha}}\,\right)\,\dr \xi,
$$
where 
\beq
C(\xi;\nu,\gamma)
:=\frac{1}{\pi}\frac{\gamma}{\left\{\,(\xi-\nu)^{\,2}+\gamma^{\,2}\right\}},
\qquad 
(\nu,\gamma)\in\rmH
\label{Cauchy-distribution}
\eeq 
with 
$\rmH:=\mbbR\times \mbbR_{>0}\subset \mbbR^{\,2}$.
\end{Proposition}
Since it is known that a mixing property leads to an ergodic property 
in general, it follows from this Proposition that 
the dynamical system associated with $F_{\,\alpha}$ has an ergodic property. 

The function $C(\xi;\nu,\gamma)$ 
is known as the Cauchy distribution,  
where $\gamma$ and $\nu$ are referred to as  
the scale parameter and the location parameter, respectively. 
This $C(\xi;\nu,\gamma)$ belongs to the stable 
distribution family, and $C(\xi;0,1)$ 
is referred to as the standard Cauchy distribution. 
The space $\rmH$ is used as a 
parameter space for expressing Cauchy distributions, and is referred to as 
the upper half-plane. 

%
\section{Parameter maps from Perron-Frobenius equations}
\label{section-parameter-maps}
In this section, a family of maps on $\rmH$
is exactly derived 
and its basic properties 
are discussed, where the derived maps describe dynamics of 
distribution functions of the generalized Boole transform family.  
Such maps are obtained by reducing the Perron-Frobenius equations,   
$\rho_{\,n}\mapsto\rho_{\,n+1}$ with $n\in\mbbZ$  
\beq
\rho_{n+1}(\,\xi^{\,\prime}\,)
=\sum_{\xi= F_{\,\alpha}^{-1}(\xi^{\,\prime})}
\frac{1}{\left|\frac{\dr F_{\,\alpha}}{\dr \xi}\right|}\,\rho_{\,n}(\xi),
\label{definition-Perron-Frobnenius-equation}
\eeq
where $\xi$ in the sum denotes the set of all the preimages of a given point 
$\xi^{\,\prime}$, and $\rho_{\,n}$ the distribution function. 
By analyzing \fr{definition-Perron-Frobnenius-equation}, one has 
the time-evolution of 
distribution functions of the dynamical system.

In this paper, the way of viewing dynamical systems with the Perron-Frobenius 
equations is referred to as the statistical picture. In addition, 
the way of viewing dynamical systems with $\{F_{\,\alpha}\}$ is referred to 
as the orbital picture.

One then has the following in the statistical picture.
This statement is the departure of the discussions below.  
\begin{Proposition}
(Parameter maps).  
Consider \fr{definition-Perron-Frobnenius-equation}, 
where $F_{\,\alpha}$ is given 
by \fr{definition-generalized-Boole-transform-alpha} with $0<\alpha<1$.
If $\rho_{\,n}(\xi)=C(\xi;\nu,\gamma)$, then 
$\rho_{\,n+1}(\xi^{\,\prime})=C(\xi^{\,\prime};\nu^{\,\prime},\gamma^{\,\prime})$ with 
$\xi^{\,\prime}=F_{\,\alpha}(\xi)$, 
\beq
\nu^{\,\prime}
=\cF_{\,\alpha,-}(\gamma,\nu)
:=\alpha\,\nu\frac{\gamma^{\,2}+\nu^{\,2}-1}{\gamma^{\,2}+\nu^{\,2}}
,\quad
\mbox{and}\quad
\gamma^{\,\prime}
=\cF_{\,\alpha,+}(\gamma,\nu)
:=\alpha\,\gamma\frac{\gamma^{\,2}+\nu^{\,2}+1}{\gamma^{\,2}+\nu^{\,2}}.
\label{parameter-map-alpha}
\eeq
\end{Proposition}
\begin{Proof}
Substituting $\rho_{\,n}(\xi)=C(\xi;\nu,\gamma)$ into 
\fr{definition-Perron-Frobnenius-equation},
one can complete the proof. The details of this calculation with a fixed  
$\alpha$ are as follows.

Since the number of points that give a point  
$\xi^{\,\prime}$ under the map iteration $\xi\mapsto \xi^{\,\prime}=F_{\,\alpha}(\xi)$ 
is two, the Perron-Frobenius equation is of the form 
$$
\rho_{\,n+1}\,(\xi^{\,\prime})
=\sum_{j=1,2}\frac{1}{\left|\frac{\dr \xi^{\,\prime}}{\dr \xi}
\right|_{\xi_{\,j}}}C(\xi_{\,j};\nu,\gamma),\qquad
\xi^{\,\prime}=F_{\,\alpha}(\xi_{\,j}).
$$
To have an explicit form of the equation above, one needs the preimage of a 
given point $\xi^{\,\prime}\in\cR_{\,-}$.  
The preimage is $\xi_{\,-}\cup \xi_{\,+}$, where $\xi_{\,\pm}$ are such that 
$\xi^{\,\prime}=F_{\,\alpha}(\xi_{\,+})$ and $\xi^{\,\prime}=F_{\,\alpha}(\xi_{\,-})$.  
They are obtained by solving
$$
\xi^{\,\prime}
=\alpha\,\left(\xi-\frac{1}{\xi}\right),
$$ 
for $\xi$ as 
$$
\xi_{\,+}(\xi^{\,\prime})
=\frac{1}{2\,\alpha}\left[\xi^{\,\prime}+\sqrt{\xi^{\,\prime\,2}+4\alpha^{\,2}}\,\right],
\qquad\mbox{and}\qquad 
\xi_{\,-}(\xi^{\,\prime})
=\frac{1}{2\,\alpha}\left[\xi^{\,\prime}-\sqrt{\xi^{\,\prime\,2}+4\alpha^{\,2}}\,\right].
$$
From these explicit expressions one has the following relations:  
$$
\xi_{\,-}+\xi_{\,+}
=\frac{\xi^{\,\prime}}{\alpha},\qquad\mbox{and}\qquad
\xi_{\,-}\,\xi_{\,+}
=-1,
$$
from which 
$$
\xi_{\,-}^{\,2}+\xi_{\,+}^{\,2}
=(\xi_{\,-}+\xi_{\,+})^{\,2}-2\xi_{\,-}\xi_{\,+}
=\left(\frac{\xi^{\,\prime}}{\alpha}\right)^{\,2}+2.
$$
In addition it follows from 
$$
4+\left(\frac{\xi^{\,\prime}}{\alpha}\right)^{\,2}
=\left(\xi+\frac{1}{\xi}\right)^{\,2},
$$
that 
$$
\frac{4+(\xi^{\,\prime}/\alpha)^{\,2}}{\xi^{\,2}+1}
=\frac{\xi^{\,2}+1}{\xi^{\,2}}.
$$
With these, one has 
$$
\left.\frac{\dr \xi^{\,\prime}}{\dr \xi}\right|_{\xi_{\,\pm}}
=\alpha\left.\frac{1+\xi^{\,2}}{\xi^{\,2}}\right|_{\xi_{\,\pm}}
=\alpha\frac{4+(\xi^{\,\prime}/\alpha)^{\,2}}{1+\xi_{\,\pm}^{\,2}}
=\left|\frac{\dr \xi^{\,\prime}}{\dr \xi}\right|_{\xi_{\,\pm}}.
$$
Thus the Perron-Frobenius equations read 
$$
\rho_{\,n+1}\,(\xi^{\,\prime})
=\frac{1}{4+(\xi^{\,\prime}/\alpha)^{\,2}}\frac{\gamma}{\alpha\pi}
\left(\frac{1+\xi_{\,+}^{\,2}}{\gamma^{\,2}+(\xi_{+}-\nu)^{\,2}}
+\frac{1+\xi_{\,-}^{\,2}}{\gamma^{\,2}+(\xi_{-}-\nu)^{\,2}}
\right).
$$
Then it follows that 
\beqa
\rho_{\,n+1}\,(\xi^{\,\prime})
&=&\frac{\gamma}{\alpha\pi\,\left[4+(\xi^{\,\prime}/\alpha)^{\,2}\right]}\left[
\frac{ (1+\xi_{\,+}^{\,2} ) \{ (\xi_{\,-}-\nu)^{\,2}+\gamma^{\,2} \} 
+(1+\xi_{\,-}^{\,2} ) \{ (\xi_{\,+}-\nu)^{\,2}+\gamma^{\,2} \} 
}{
\gamma^{\,4}+\gamma^{\,2}\left\{ (\xi_{\,-}-\nu)^{\,2}+(\xi_{\,+}-\nu)^{\,2} \right\}
+(\xi_{\,-}-\nu)^{\,2}(\xi_{\,+}-\nu)^{\,2} }\right]
\non\\
&=&\frac{\gamma}{\alpha\,\pi\,\left[4+(\xi^{\,\prime}/\alpha)^{\,2}\right]}
\frac{\Upsilon_{\mathrm{n}}}{\Upsilon_{\mathrm{d}}},
\non
\eeqa
where $\Upsilon_{\mathrm{n}}$ and $\Upsilon_{\mathrm{d}}$ are given by 
\beqa
\Upsilon_{\mathrm{n}}
&=&
 (1+\xi_{\,+}^{\,2} ) \{ (\xi_{\,-}-\nu)^{\,2}+\gamma^{\,2} \} 
+(1+\xi_{\,-}^{\,2} ) \{ (\xi_{\,+}-\nu)^{\,2}+\gamma^{\,2} \},
\non\\
\Upsilon_{\mathrm{d}}
&=& 
\gamma^{\,4}+\gamma^{\,2}
\left\{ (\xi_{\,-}-\nu)^{\,2}+(\xi_{\,+}-\nu)^{\,2} \right\}
+(\xi_{\,-}-\nu)^{\,2}(\xi_{\,+}-\nu)^{\,2},
\non
\eeqa
respectively. 
The expression of $\Upsilon_{\mathrm{d}}$ and that of 
$\Upsilon_{\mathrm{d}}$ reduce to  
\beqa
\Upsilon_{\mathrm{n}}
&=&(\gamma^{\,2}+\nu^{\,2}+1)\left[\,
4+\left(\frac{\xi^{\,\prime}}{\alpha}\right)^{\,2}
\,\right],
\non\\
\Upsilon_{\mathrm{d}}
&=&(\gamma^{\,2}+\nu^{\,2})\left(\frac{\xi^{\,\prime}}{\alpha}\right)^{\,2}
-2\nu(\gamma^{\,2}+\nu^{\,2}-1)\frac{\xi^{\,\prime}}{\alpha}
+\gamma^{\,4}+2\gamma^{\,2}(\nu^{\,2}+1)+(\nu^{\,2}-1)^{\,2}.
\non
\eeqa
The expression of $\Upsilon_{\mathrm{d}}$ reduces further by introducing
$$
\nu^{\,\prime}
=\alpha\frac{\nu\,(\gamma^{\,2}+\nu^{\,2}-1)}{\gamma^{\,2}+\nu^{\,2}},
$$
as 
$$
\Upsilon_{\mathrm{d}}
=\frac{\gamma^{\,2}+\nu^{\,2}}{\alpha^{\,2}}
(\xi^{\,\prime}-\nu^{\,\prime})^{\,2}+4\gamma^{\,2}\left[
1+\frac{\gamma^{\,2}+\nu^{\,2}}{4\alpha^{\,2}}\left(\frac{\nu^{\,\prime}}{\nu}\right)^{\,2}
\right],
$$
where the relation 
$$
\gamma^{\,4}+2\gamma^{\,2}(\nu^{\,2}+1)
=4\gamma^{\,2}\left[\,1+
\frac{\gamma^{\,2}+\nu^{\,2}}{4\alpha^{\,2}}
\left(\frac{\nu^{\,\prime}}{\nu}\right)^{\,2}\,\right]
-\frac{\gamma^{\,2}}{\gamma^{\,2}+\nu^{\,2}}+\gamma^{\,2}\nu^{\,2},
$$
has been used. 
Combining $\Upsilon_{\mathrm{n}}$ and $\Upsilon_{\mathrm{d}}$, one has 
$$
\rho_{\,n+1}\,(\xi^{\,\prime})
=\frac{\gamma}{\pi}
\frac{1}{\alpha\left[4+\left(\frac{\xi^{\,\prime}}{\alpha}\right)^{\,2}\right]}
\frac{\left[4+\left(\frac{\xi^{\,\prime}}{\alpha}\right)^{\,2}\right]\left(
\gamma^{\,2}+\nu^{\,2}+1\right)}{\left[\frac{\gamma^{\,2}+\nu^{\,2}}{\alpha^{\,2}}
(\xi^{\,\prime}-\nu^{\,\prime})^{\,2}+4\gamma^{\,2}\left\{
1+\frac{\gamma^{\,2}+\nu^{\,2}}{4\alpha^{\,2}}\left(\frac{\nu^{\,\prime}}{\nu}\right)^{\,2}
\right\}\right]},
$$
from which 
$$
\rho_{\,n+1}(\xi^{\,\prime})
=\frac{1}{\pi}
\frac{\alpha\frac{\gamma(\gamma^{\,2}+\nu^{\,2}+1)}
{\gamma^{\,2}+\nu^{\,2}}}{(\xi^{\,\prime}-\nu^{\,\prime})^{\,2}+\left[
\left(\gamma\frac{\nu^{\,\prime}}{\nu}\right)^{\,2}+\frac{4\alpha^{\,2}\gamma^{\,2}}{\gamma^{\,2}+\nu^{\,2}}\right]}.
$$
The term in the bracket $\left[\cdots\right]$ above can be written by 
introducing 
$$
\gamma^{\,\prime}
=\alpha\frac{\gamma\,(\gamma^{\,2}+\nu^{\,2}+1)}{\gamma^{\,2}+\nu^{\,2}},
$$
as 
$$
\left(\gamma\frac{\nu^{\,\prime}}{\nu}\right)^{\,2}
+\frac{4\alpha^{\,2}\gamma^{\,2}}{\gamma^{\,2}+\nu^{\,2}}
=(\,\gamma^{\,\prime}\,)^{\,2}.
$$
Thus, one arrives at 
$$
\rho_{\,n+1}\,(\xi^{\,\prime})
=\frac{1}{\pi}\left[\frac{\gamma^{\,\prime}}{(\xi^{\,\prime}-\nu^{\,\prime})^{\,2}+(\gamma^{\,\prime})^{\,2}}\right],
$$
from which one concludes that 
$\rho_{\,n+1}\,(\xi^{\,\prime})=C(\xi^{\,\prime};\nu^{\,\prime},\gamma^{\,\prime})$.
\qed
\end{Proof}
In this paper,  
the family of maps 
$\rmH\to\rmH,  (\nu,\gamma)\mapsto(\nu^{\,\prime},\gamma^{\,\prime})$ 
is referred to as the parameter maps or the parameter map family, 
and $\rmH$ the phase space of the parameter maps. 
For a fixed $\alpha$, 
the parameter map is denoted $\cF_{\,\alpha}$. 
When one needs to avoid the infinite points in $\rmH$,  
one restricts $\rmH$ to a subspace of $\rmH$.  
Remarks on this family of parameter maps are listed below. 
\begin{enumerate}
\item 
The case of $\alpha=1/2$ was considered in Ref.\,\cite{Shintani-Umeno2017}, 
and the parameter map for $\alpha=1/2$ was obtained. 
That previously derived map is consistent with \fr{parameter-map-alpha}. 

\item
When $\gamma^{\,2},\nu^{\,2}\gg 1$, the parameter maps  
\fr{parameter-map-alpha}
are approximately written as  
$$
\gamma_{\,n+1}
=\alpha\,\gamma_{\,n},\qquad \mbox{and}\qquad
\nu_{\,n+1}
=\alpha\,\nu_{\,n}.
$$
The  solution set is immediately obtained as 
$$
\gamma_{\,n}
=\alpha^{\,n}\,\gamma_{\,0},\qquad\mbox{and}\qquad 
\nu_{\,n}
=\alpha^{\,n}\,\nu_{\,0}.
$$
\item
There exist at least two invariant manifolds when the phase space is extended. 
They are 
$$
\rmH_{\,\nu=0}
:=\{\,(\nu,\gamma)\,\in\rmH\,|\,\nu=0\,\},\qquad\mbox{and}\qquad 
\ol{\rmH}_{\,\gamma=0}
:=\{\,(\nu,\gamma)\,\in\ol{\rmH}\,|\,\gamma=0\,\},\qquad 
$$ 
where $\ol{\rmH}:=\mbbR\times\mbbR_{\geq 0}\subset\mbbR^{\,2}$.
The dynamical system on each invariant manifold is 
$$
\rmH_{\,\nu=0}\, :\, \gamma^{\,\prime}
=G_{\,\alpha}(\gamma),\qquad\mbox{and}\qquad  
\ol{\rmH}_{\,\gamma=0}\,:\,
\nu^{\,\prime}
=F_{\,\alpha}(\nu),
$$  
where $F_{\,\alpha}$ has been defined in 
\fr{definition-generalized-Boole-transform-alpha} and 
$G_{\,\alpha}:\cR_{\,+}\to\cR_{\,+}$ with $\cR_{\,+}=\mbbR\setminus\cR_{\,+}^{\,\infty}$
is such that  
\beq
G_{\,\alpha}(\gamma)
:=\alpha\left(\gamma+\frac{1}{\gamma}\right).
\label{G-alpha}
\eeq
Here the set $\cR_{\,+}^{\,\infty}$ is a collection of points of $\mbbR$ so that 
$G_{\,\alpha}(\gamma)$ is finite for all points of $\cR_{\,+}$.
On the boundary of the extended 
phase space $\ol{\rmH}$, one can 
consider Dirac's delta function as the limiting distribution function,   
$$
\delta(\xi-\nu)
=C(\xi;\nu,0).
$$
The dynamics of $\nu$ takes place on $\ol{\rmH}_{\,\gamma=0}$ and is exactly the 
same as that of 
the orbital picture.
Note that the function $G_{\,1/2}$ is the same as $G$ introduced 
in Ref.\,\cite{Umeno2016}.
\item
A fixed point for the map with a fixed $\alpha$ is  
$(\,\ol{\nu},\ol{\gamma}\,)\in\rmH$, where 
\beq
\ol{\nu}
=0\qquad\mbox{and}\qquad 
\ol{\gamma}
=\sqrt{\frac{\alpha}{1-\alpha}}.
\label{fixed-point-parameter-map}
\eeq
This fixed point is unique for a fixed $\alpha$. 
\begin{Proof}
To prove this, the statement is split into the following two.  
\begin{enumerate}
\item
There is no fixed point on the set 
$\{(\nu,\gamma)|\nu\neq 0,\gamma>0\}\subset \rmH$.
\item
The fixed point on $\rmH$ is given by \fr{fixed-point-parameter-map}, 
and is unique.
\end{enumerate}
As a notational convenience introduce $\ol{A}=\ol{\nu}^{\,2}+\ol{\gamma}^{\,2}$. 

(Proof of (a)\,) : Assume that there exists a fixed point 
$(\ol{\nu},\ol{\gamma})$ with $\ol{\nu}\neq 0$ and $\ol{\gamma}>0$. 
Then the fixed point are given by the solutions to 
$$
(1-\alpha)\,\ol{A}
=-\,\alpha,\quad\mbox{and}\quad
(1-\alpha)\,\ol{A}
=\,\alpha.
$$
Taking into account $0<\alpha<1$, one has that $\ol{A}=0$, from which 
$(\ol{\nu},\ol{\gamma})=(0,0)$. This is in contradiction to 
the assumption on $(\ol{\nu},\ol{\gamma})$.
This completes the proof of (a).

(Proof of (b)\,) :  Taking into account (a), one looks for fixed points on 
$\rmH_{\,\nu=0}=\{(\nu,\gamma)|\nu=0,\gamma>0\}$.
The equation for determining fixed points on $\rmH_{\,\nu=0}$ is derived 
by substituting $\nu=0$ into  
\fr{parameter-map-alpha} as 
$$
(1-\alpha)\ol{\gamma}^{\,2}
=\alpha.
$$
From this equation, there exists 
only one solution on $\rmH$ as $\ol{\gamma}=\sqrt{\alpha/(1-\alpha)}$ .
\qed
\end{Proof}
When $\alpha=1/2$, 
one has $(\ol{\nu},\ol{\gamma})=(0,1)$ (see Ref.\,\cite{Shintani-Umeno2017}). 
This fixed point corresponds to the standard Cauchy distribution.
\item
The linearized map 
around $(\,\ol{\nu},\ol{\gamma}\,)$, denoted 
$(\delta\nu,\delta\gamma)\mapsto (\delta\nu^{\,\prime},\delta\gamma^{\,\prime})$, 
is obtained from 
$$
\delta\gamma^{\,\prime}
=\left(\frac{\partial \cF_{\,\alpha,+}}{\partial \gamma}
\right)_{(\ol{\nu},\ol{\gamma})}
\delta\gamma
+\left(\frac{\partial \cF_{\,\alpha,+}}{\partial \nu}\right)_{(\ol{\nu},\ol{\gamma})}\delta\nu,
\qquad
\delta\nu^{\,\prime}
=\left(\frac{\partial \cF_{\,\alpha,-}}{\partial \gamma}\right)_{(\ol{\nu},\ol{\gamma})}
\delta\gamma
+\left(\frac{\partial \cF_{\,\alpha,-}}{\partial \nu}\right)_{(\ol{\nu},\ol{\gamma})}\delta\nu,
$$
as 
$$
\delta\gamma^{\,\prime}
=(2\alpha-1)\,\delta\gamma,\qquad\mbox{and}\qquad
\delta\nu^{\,\prime}
=(2\alpha-1)\,\delta\nu.
$$
This shows that 
this fixed point is linearly stable for $0<\alpha<1$ except for $\alpha=1/2$.  
Notice that the linear stability analysis fails when $\alpha=1/2$, and 
sequences converge in a quadratic order \cite{Nakamura2001}.  
\item
An estimate of the convergence of sequences on $\rmH_{\,\nu=0}$ is given  
as follows. For $n\geq 2$, 
\beqa
|\,\gamma_{\,n+1}-\ol{\gamma}\,|
&\leq& \,\alpha|\,\gamma_{\,n}-\ol{\gamma}\,|,\qquad 
\mbox{for}\quad\frac{1}{2}\leq \alpha<1
\non\\
|\,\gamma_{\,n+1}-\ol{\gamma}\,|
&\leq& \,(1-\alpha)\,|\,\gamma_{\,n}-\ol{\gamma}\,|,\qquad
\mbox{for}\quad 0<\alpha \leq \frac{1}{2}
\non
\eeqa
where an initial point is 
 specified with $n=0$.
\item
Fix $\alpha$. Then let $\cF_{\,\alpha}:\rmH\to\rmH$,  
$(\nu,\gamma)\mapsto(\nu^{\,\prime},\gamma^{\,\prime})$ be the parameter map, 
and $\cI:\rmH\to\rmH$,$(\nu,\gamma)\mapsto (-\nu,\gamma)$ 
the reflection operator.  Then it follows from 
$$
\cI\cF_{\,\alpha}(\nu,\gamma)
=\cI(\nu^{\,\prime},\gamma^{\,\prime})
=(-\nu^{\,\prime},\gamma^{\,\prime}),
\qquad\mbox{and}\qquad 
\cF_{\,\alpha}\cI(\nu,\gamma)
=\cF_{\,\alpha}\,(-\nu,\gamma)
=(-\nu^{\,\prime},\gamma^{\,\prime})
$$
that $\cF_{\,\alpha}$ and $\cI$ are commute,  
$\cF_{\,\alpha}\cI=\cI\cF_{\,\alpha}$.
\end{enumerate}

When introducing a set of complex variables, one can formally rewrite 
the parameter maps $\{\cF_{\,\alpha}\}$ with $\{F_{\,\alpha}\}$ 
as follows. 
\begin{Lemma}
\label{goto-lemma-1}
Fix $\alpha$. Then let $(\nu,\gamma)$ and 
$(\nu^{\,\prime},\gamma^{\,\prime})$ be points of $\rmH$ that satisfy 
\fr{parameter-map-alpha}. Define the complex variables 
$s,w,s^{\,\prime},w^{\,\prime}\in\mbbC$ to be   
$$
s:=\nu-\mi \gamma,\quad 
w:=\nu+\mi \gamma,\quad 
s^{\,\prime}:=\nu^{\,\prime}-\mi \gamma^{\,\prime},\quad 
w^{\,\prime}:=\nu^{\,\prime}+\mi \gamma^{\,\prime},\quad
\mi:=\sqrt{-1}.
$$ 
Then the map $(s,w)\mapsto(s^{\,\prime},w^{\,\prime})$ under 
$(\nu,\gamma)\mapsto(\nu^{\,\prime},\gamma^{\,\prime})$ is the following 
set of maps:  
$$
s^{\,\prime}
=F_{\,\alpha}(s),\qquad\mbox{and}\qquad  
w^{\,\prime}
=F_{\,\alpha}(w).
$$
\end{Lemma}
\begin{Proof}
One completes the proof by  
substituting $s,w,s^{\,\prime},w^{\,\prime}$ defined above into  
\fr{parameter-map-alpha}. The details are as follows. Introducing 
$A=\nu^{\,2}+\gamma^{\,2}=sw$, one has
$$
\nu^{\,\prime}
=\alpha\nu\left(1-\frac{1}{sw}\right),\quad\mbox{and}\quad
\gamma^{\,\prime}
=\alpha\gamma\left(1+\frac{1}{sw}\right).
$$ 
With these, it follows that 
$$
s^{\,\prime}
=\nu^{\,\prime}-\mi\gamma^{\,\prime}
=\alpha\left(s-\frac{1}{s}\right),\quad\mbox{and}\quad
w^{\,\prime}
=\nu^{\,\prime}+\mi\gamma^{\,\prime}
=\alpha\left(w-\frac{1}{w}\right).
$$
\qed
\end{Proof}
Similar to the lemma above, one has the following. 
\begin{Lemma}
\label{goto-lemma-2}
Consider Lemma\,\ref{goto-lemma-1}. 
Define $\check{s},\check{w},\check{s}^{\,\prime},\check{w}^{\,\prime}\in\mbbC$ to 
be  
$$
\check{s}
:=\mi s
=\gamma+\mi\nu,\quad
\check{w}
:=\mi w
=-\,\gamma+\mi\nu,\quad 
\check{s}^{\,\prime}
:=\mi s^{\,\prime}
=\gamma^{\,\prime}+\mi\nu^{\,\prime},\quad
\check{w}^{\,\prime}
:=\mi w^{\,\prime}
=-\,\gamma^{\,\prime}+\mi\nu^{\,\prime}.
$$
Then the map 
$(\check{s},\check{w})\mapsto(\check{s}^{\,\prime},\check{w}^{\,\prime}) $ 
under $(\nu,\gamma)\mapsto(\nu^{\,\prime},\gamma^{\,\prime})$ is the following set 
of maps:   
$$
\check{s}^{\,\prime}
=\alpha\left(\check{s}+\frac{1}{\check{s}}\right)
=G_{\,\alpha}(\check{s}),\qquad \mbox{and}\qquad 
\check{w}^{\,\prime}
=\alpha\left(\check{w}+\frac{1}{\check{w}}\right)
=G_{\,\alpha}(\check{w}),
$$
where $G_{\,\alpha}(\check{s})=\alpha(\check{s}+1/\check{s})$ has been 
defined by \fr{G-alpha}.
\end{Lemma}
\begin{Proof}
A way to prove this is analogous to that of Lemma\,\ref{goto-lemma-1}. 
\qed
\end{Proof}

The role of the complex variables $s,w,s^{\,\prime},w^{\,\prime}$ is 
to decompose \fr{parameter-map-alpha} into a set of $F_{\,\alpha}$  
being extended for complex-variables. 
Each decomposed map is used in the orbital picture. 
Thus Lemma\,\ref{goto-lemma-1} gives a relation between 
the statistical picture and orbital one. 
The next one  plays the same role.
\begin{Lemma}
\label{goto-lemma-3}
Let $\cF_{\,\alpha}:\rmH\to\rmH$ be the the parameter map 
defined by \fr{parameter-map-alpha}:  
$$
\nu^{\,\prime}
:=\cF_{\,\alpha,-}(\gamma,\nu)
,\qquad
\gamma^{\,\prime}
:=\cF_{\,\alpha,+}(\gamma,\nu). 
$$ 
In addition, let $(\xi^{\,1},\xi^{\,2})\in\cR_{\,-}\times\cR_{\,-}$ be a point.  
Then, introducing the complex variables 
$$
\wt{\xi}
:=\xi^{\,1}+\mi \xi^{\,2}\in\mbbC,\qquad\mbox{and}\qquad
\wt{\xi}^{\,\prime}
:=F_{\,\alpha}(\,\wt{\xi}\,)
=\xi^{\,1\,\prime}+\mi \xi^{\,2\,\prime}\in\mbbC, 
$$
with $\xi^{\,1\,\prime}=\Ree(\wt{\xi}^{\,\prime})$ and 
$\xi^{\,2\,\prime}=\Imm(\wt{\xi}^{\,\prime})$, 
one has 
$\wt{\xi}^{\,\prime}
=\cF_{\,\alpha,-}(\xi^{\,1},\xi^{\,2})
+\mi\,\cF_{\,\alpha,+}(\xi^{\,1},\xi^{\,2})$ : 
$$
\xi^{\,1\,\prime}
=\cF_{\,\alpha,-}(\xi^{\,1},\xi^{\,2}),\quad\mbox{and}\quad
\xi^{\,2\,\prime}
=\cF_{\,\alpha,+}(\xi^{\,1},\xi^{\,2}).
$$
\end{Lemma}
\begin{Proof}
Calculating $F_{\,\alpha}(\,\wt{\xi}\,)$, one can complete the proof.   
Substituting $\wt{\xi}=\xi^{\,1}+\mi \xi^{\,2}$ into 
$F_{\,\alpha}(\,\wt{\xi}\,)$, one has 
\beqa
F_{\,\alpha}(\,\wt{\xi}\,)
&=&\alpha\left(\wt{\xi}-\frac{1}{\wt{\xi}}\right)
=\alpha\left(\xi^{\,1}+\mi \xi^{\,2}-\frac{1}{\xi^{\,1}+\mi \xi^{\,2}}\right)
\non\\
&=&\alpha \xi^{\,1}\frac{(\xi^{\,1})^{\,2}+(\xi^{\,2})^{\,2}-1}{(\xi^{\,1})^{\,2}+(\xi^{\,2})^{\,2}}+\mi 
\alpha \xi^{\,2}\frac{(\xi^{\,1})^{\,2}+(\xi^{\,2})^{\,2}+1}{(\xi^{\,1})^{\,2}+(\xi^{\,2})^{\,2}}
=\cF_{\,\alpha,-}(\xi^{\,1},\xi^{\,2})+\mi 
\cF_{\,\alpha,+}(\xi^{\,1},\xi^{\,2}).
\non
\eeqa
On the other hand, the left hand side of the equation above 
is $F_{\,\alpha}(\wt{\xi})=\wt{\xi}^{\,\prime}=\xi^{\,1\,\prime}+\mi \xi^{\,2\,\prime}$.
\qed
\end{Proof}
Due to this lemma, the maps $\{F_{\alpha}\}$ used in 
the orbital picture 
can be written with $\{\cF_{\,\alpha}\}$ used in 
the statistical picture.  

Combining Lemmas\,\ref{goto-lemma-1} and \ref{goto-lemma-3},  
one has the following. 
\begin{Thm}
(Relation between statistical picture and orbital one):  
Fix $\alpha$. Then the parameter map $\cF_{\,\alpha}$   
can be written in terms of $F_{\,\alpha}$ with some complex variables. 
On the other hand, the map $\xi^{\,\prime}=F_{\,\alpha}(\xi)$ in the orbital 
picture is written in terms of the parameter map with some complex variables. 
\end{Thm}

The reason why the finite-dimensional parameter maps have successfully 
been obtained is to choose 
a particular family of distribution functions  
for the Perron-Frobenius equations. In our case,  
the Cauchy distribution with a parameter set is chosen, and 
it has turned out that the 
distribution function with $(\nu,\gamma)$ 
approaches to the one with $(\ol{\nu},\ol{\gamma})$ after 
a long-time evolution if an
 initial parameter set is close enough to 
$(\ol{\nu},\ol{\gamma})$. 
Also, notice that the dimension of the phase space $\rmH$ 
of the  parameter map family $\{\cF_{\,\alpha}\}$ is the number of the 
parameters of the Cauchy distribution.   
This dimension of the paramter map family 
 is not directly related to the dimension of 
the phase space of $\{F_{\,\alpha}\}$. 
One can generalize this notice as follows. 
Consider the case where a map in an $n$-dimensional phase space is solvable,
and its invariant measure is written with a distribution function having 
$k$-parameters. 
Assume that a parameter map is obtained 
by reducing the Perron-Frobenius equation.
Then the dimension of the phase space of a parameter map is $k$, and 
$n$ is not directly related to  $k$ in general.
 
\section{Information and symplectic geometric characterizations}
In Section \ref{section-parameter-maps}, the parameter maps have been derived and some relations
between the orbital picture and statistical one have been obtained. 

In this section,  
dynamics of the parameter maps in 
the phase space is characterized with 
information geometry and symplectic geometry. 
Since information geometry gives differential geometric tools for analyzing 
distribution functions on parameter spaces (or phase spaces),  
it is expected that the parameter maps can be characterized with such tools. 
Also, since it has been known how a symplectic structure is induced from 
a statistical manifold, one can apply symplectic 
geometry to the parameter maps.  
In particular,  
it is shown that the parameter maps are conformal 
on a Riemannian manifold and that 
the parameter maps are written with canonical coordinates. 
To this end, some connections and manifolds are introduced first. 

\subsection{Information geometric characterization of  parameter maps}

The following connection plays roles in information geometry.
\begin{Def}
\label{definition-dual-connection}
(Dual connection, \cite{AN}).    
Let $(\cM,g)$ be a (pseudo) Riemannian manifold, and $\nabla$ a connection.
The dual connection $\nabla^{\,*}$ associated with $g$ 
is a connection that satisfies 
$X(g(Y,Z))=g(\nabla_{X}Y,Z)+g(Y,\nabla_{X}^{\,*}Z)$ for all $X,Y,Z\in T\cM$
\end{Def}
It can be verified that the dual connection is determined uniquely from 
$\nabla$ and $g$, and that $(\nabla^{\,*})^{\,*}=\nabla$.  

The following is our definition of statistical manifold.
\begin{Def}
(Statistical manifold, \cite{Takano2006}).    
Let $(\cM,g)$ be a (pseudo) Riemannian manifold, $\nabla$ a connection, 
and $\nabla^{\,*}$ the dual connection associated with $g$. 
If $\nabla$ and $\nabla^{\,*}$ are torsion-free connections, 
then the triplet $(\cM,g,\nabla)$ is referred to as 
the statistical manifold.  
\end{Def} 
Note that any flatness condition is not imposed in the definition above.
The torsion-free condition is written as 
$$
T^{\,\nabla}(X,Y)
=\nabla_{\,X}Y-\nabla_{\,Y}X-[X,Y]
=0,\qquad \mbox{for all $X,Y\in T\cM$} 
$$ 
with $[X,Y]=XY-YX$. Here  
$T^{\nabla}$ is referred to as torsion tensor.

A role of the Levi-Civita connection on statistical manifolds is as follows.
\begin{Lemma}
\label{fact-statistical-manifold-Levi-Civita}
Let $(\cM,g,\nabla)$ is a statistical manifold. 
Then, 
$$
\nabla=\nabla^{\,*}
\qquad\Longleftrightarrow\qquad 
\nabla\quad \mbox{is the Levi-Civita connection}
$$
\end{Lemma}
\begin{Proof}
(Proof of $\Rightarrow$) : From the definition of statistical manifold, 
one has  
that $T^{\nabla}=0$. Then from the assumption, one has that  
$X(g(Y,Z))=g(\nabla_{\,X}Y,Z)+g(Y,\nabla_{\,X}Z)$, which is equivalent to 
$\nabla g=0$.  
These two conditions 
guarantee that $\nabla$ is the Levi-Civita connection.\\
(Proof of $\Leftarrow$) : 
Since $\nabla$ is the Levi-Civita connection, one has 
$\nabla g=0$ :  
$$
X(g(Y,Z))=g(\nabla_{\,X}Y,Z)+g(Y,\nabla_{\,X}Z).
$$
Comparing this with the definition of dual connection
Definition.\,\ref{definition-dual-connection}, one has that $\nabla=\nabla^{\,*}$. 
\qed 
\end{Proof}

In the standard information geometry, 
parametric distribution functions are considered. 
They form the following set.  
\begin{Def}
(Statistical model, \cite{AN}).    
Let $\cS$ be a set of probability distribution functions 
that are parameterized by $n$
real-valued variables $\zeta=(\zeta^{\,1},\ldots,\zeta^{\,n})$ so that 
$$
\cS=\{\,p_{\,\zeta}=p(\xi;\zeta)\,|\,
\zeta=(\zeta^{\,1},\ldots,\zeta^{\,n})\in \cZ\,\,\},
$$ 
where $\cZ$ is a subset of $\mbbR^{\,n}$ and 
the map $\zeta\mapsto p_{\zeta}$ is injective.
This $\cS$ is referred to as an $n$-dimensional statistical model.
\end{Def}
It is assumed that the space of 
parameters for parametric distribution functions
form a manifold. 
\begin{Postulate}
Any $n$-dimensional statistical model forms an $n$-dimensional manifold.
\end{Postulate}

For a given probability distribution function,  
the Fisher metric tensor field is defined below. 
\begin{Def}
(Fisher information matrix, Fisher metric tensor field, \cite{AN}).   
Let $\cS=\{p_{\,\zeta}\,|\,\zeta\in \cZ\,\}$ 
be an $n$-dimensional statistical model. 
Given a point $\zeta$, the $n\times n$ matrix $\{g_{\,ab}^{\,\F}\}$ whose 
elements are defined by 
$$
g_{\,ab}^{\,\F}(\zeta)
:=\int \frac{\partial\ln p(\xi;\zeta)}{\partial \zeta^{\,a}}
\frac{\partial\ln p(\xi;\zeta)}{\partial \zeta^{\,b}}
p(\xi;\zeta)\dr \xi
$$
is referred to as the Fisher information matrix. 
In addition, the metric tensor field 
$$
g^{\,\F}(\zeta)
:=g_{\,ab}^{\,\F}(\zeta)\,\dr \zeta^{\,a}\otimes \dr\zeta^{\,b}, 
$$
is referred to as 
the Fisher metric tensor field. 
\end{Def}
The following is a relevant example for the parameter maps discussed in 
Section\,\ref{section-parameter-maps}.
\begin{Example}
(Fisher metric tensor field for the Cauchy distribution).  
For the Cauchy distribution 
$p_{\zeta}=p(\xi;\zeta)=C(\xi;\nu,\gamma)$  
 given in \fr{Cauchy-distribution} with $\zeta=(\nu,\gamma)$,  
the explicit form of the Fisher metric tensor field is calculated as follows. 
It can be shown that 
$g_{\,\nu\nu}^{\,\F}=g_{\,\gamma\gamma}^{\,\F}=1/(2\gamma^{\,2})$ 
and $g_{\nu\gamma}^{\,\F}=g_{\gamma\nu}^{\,\F}=0$, from 
which
\beq
g^{\,\F}(\nu,\gamma)
=\frac{\dr\nu\otimes\dr \nu+\dr\gamma\otimes\dr\gamma}{2\,\gamma^{\,2}}.
\label{Fisher-metric-Cauchy-nu-gamma}
\eeq
The Killing vector fields $\{\,K_{\,a}\,\}\in T\rmH$, 
defined to be vector fields 
satisfying $\cL_{\,K_{\,a}}g^{\,\F}=0$ with $\cL_{\,K_{\,a}}$ 
being the Lie derivative along $K_{\,a}$, 
are 
found as $\{\,K_{\,1},K_{\,2},K_{\,3}\,\}$ where 
\beq
K_{\,1}
=(\nu^{\,2}-\gamma^{\,2})\frac{\partial}{\partial \nu}
+2\,\nu\,\gamma\frac{\partial}{\partial \gamma},\qquad
K_{\,2}
=\nu\frac{\partial}{\partial \nu}+\gamma\frac{\partial}{\partial \gamma},\quad
K_{\,3}
=\frac{\partial}{\partial \nu}.
\label{Killing-Poincare-metric}
\eeq  
In addition, the Gaussian curvature for $\rmH$ is known to be negative.
\end{Example}

Consider an $n$-dimensional (pseudo) Riemannian manifold $(\cM,g)$.   
If the number of Killing vector fields on $\cM$ is $n(n+1)/2$,  
then the manifold is referred to as a maximally symmetric space. 
Thus the manifold $\rmH$ is a maximally symmetric space. 

The following is our main claim in this subsection and it 
characterizes our parameter maps with the Fisher metric tensor field.
\begin{Proposition}
(Parameter map as a conformal map).   
Let $(\rmH,g^{\,\F})$ be the two-dimensional Riemannian manifold where 
$\rmH$ is the phase space of 
the parameter maps (the upper half-plane),  
and $g^{\,\F}$ the Fisher metric 
tensor field given by \fr{Fisher-metric-Cauchy-nu-gamma}. 
Then the parameter maps are conformal\cite{Nakahara} 
if $(\gamma^{\,2}+\nu^{\,2})^{\,2}+1+2(\nu^{\,2}-\gamma^{\,2})>0$ 
in the sense that 
$$
g^{\,\F\,\prime}
:=\frac{\dr\nu^{\,\prime}\otimes\dr\nu^{\,\prime}
+\dr\gamma^{\,\prime}\otimes\dr \gamma^{\,\prime}}{2\,(\gamma^{\,\prime})^{\,2}}
=\frac{(\gamma^{\,2}+\nu^{\,2})^{\,2}+1+2(\nu^{\,2}-\gamma^{\,2})}
{\gamma^{\,2}+\nu^{\,2}+1}g^{\,\F},
$$
where $\nu^{\,\prime}=\cF_{\,\alpha,-}(\gamma,\nu)$ and 
$\gamma^{\,\prime}=\cF_{\,\alpha,+}(\gamma,\nu)$.
\end{Proposition}
\begin{Proof}
One can complete this proof by the following straightforward calculations. 
As a  notational convenience introduce $A=\nu^{\,2}+\gamma^{\,2}$. 
Then one starts 
with  
$$
\dr\nu^{\,\prime}
=\alpha\left[\left(1-\frac{1}{A}\right)\dr\nu+\frac{\nu}{A^{^2}}\dr A\right]
,\quad\mbox{and}\quad
\dr\gamma^{\,\prime}
=\alpha\left[\left(1+\frac{1}{A}\right)\dr\gamma-\frac{\nu}{A^{^2}}\dr A\right].
$$
After some tedious calculations, one has from 
$$
\left(1-\frac{1}{A}\right)^{\,2}+\frac{4\nu^{\,2}}{A^{\,2}}
=\left(1+\frac{1}{A}\right)^{\,2}-\frac{4\gamma^{\,2}}{A^{\,2}},
$$
that  
$$
\frac{1}{\alpha^{\,2}}\left(\,
\dr\nu^{\,\prime}\otimes\dr\nu^{\prime}
+\dr\gamma^{\,\prime}\otimes\dr\gamma^{\prime}
\,\right)
=\left[\,\left(1+\frac{1}{A}\right)^{\,2}-\frac{4\gamma^{\,2}}{A^{\,2}}
\,\right]\,\left(\,\dr\nu\otimes\dr\nu+\dr\gamma\otimes\dr\gamma\,\right).
$$
With this and $\gamma^{\,\prime}=\alpha\gamma\,\,(\,1+1/A\,)$, one has
that 
$$
\frac{\dr\nu^{\,\prime}\otimes\dr\nu^{\prime}
+\dr\gamma^{\,\prime}\otimes\dr\gamma^{\prime}
}{2\,(\,\gamma^{\,\prime}\,)^{\,2}}
=\left[1-\frac{4\gamma^{\,2}}{(1+A)^{\,2}}\right]
\frac{\dr\nu\otimes\dr\nu+\dr\gamma\otimes\dr\gamma}{2\gamma^{\,2}}
=\left[1-\frac{4\gamma^{\,2}}{(1+A)^{\,2}}\right]
g^{\,\F}.
$$
Substituting $A=\nu^{\,2}+\gamma^{\,2}$ into the equation above, 
one completes the proof.
\qed
\end{Proof}
Notice that the $\alpha$ does not appear in this Proposition.

\subsection{Symplectic information geometric characterization of parameter maps}

Since the dimension of the phase space of our parameter maps is even and 
the complex variables play  roles as seen in 
Section\,\ref{section-parameter-maps}, 
one is interested in almost 
complex manifolds and its sub-classes. The definition of almost 
complex manifold is as follows. 
\begin{Def}
(Almost complex structure and almost complex manifold, \cite{Kobayashi1969,Takano2006}). 
Let $\cM$ be a manifold. Almost complex structure is a type $(1,1)$  tensor 
field such that $J\circ J=-\,\Id$ with $\Id$ being the identical operator.
In addition, the pair $(\cM,J)$ is referred to as an almost complex manifold. 
\end{Def}
It is known that the dimension of almost complex manifolds is even and such 
manifolds are orientable. 

Some classes of almost complex manifolds are known as follows. 
\begin{Def}
(Almost Hermite manifold, \cite{Kobayashi1969,Takano2006}).     
Let $(\cM,g)$ be a (pseudo) Riemannian manifold, 
and $J$ an almost complex structure. If 
$g(JX,JY)=g(X,Y)$ is satisfied for all $X,Y\in T\cM$, then 
the triplet $(\cM,g,J)$ is referred to as an almost Hermite manifold.
\end{Def}
\begin{Def}
(Almost K\"ahler manifold, \cite{Kobayashi1969}).  
Let $(\cM,g,J)$ be an almost Hermite manifold, and $\omega$ the two-form  
defined by 
\beq
\omega(X,Y)
=g(JX,Y),
\label{Kahler-form}
\eeq
for all $X,Y\in T\cM$. If $\dr\omega=0$, then $(\cM,g,J)$ is referred to as 
an almost K\"ahler manifold.
\end{Def}
Since Killing vector fields on Riemannian manifolds play roles, 
one is interested in roles of  Killing vector fields on almost 
K\"ahler manifolds. 
\begin{Lemma}
\label{fact-almost-Kahler-Killing}
Let $(\cM,g,J)$ be an almost K\"ahler manifold, and $K$ a Killing vector 
field ($\cL_{\,K}g=0$). If  
$\cL_{\,K}J=0$, then $\cL_{\,K}\omega=0$. 
\end{Lemma}
\begin{Proof}
Applying $\cL_{\,K}$ on the both side of  
\fr{Kahler-form}, one completes the proof.
\qed
\end{Proof}

To discuss K\"ahler manifolds, one needs the following. 
\begin{Def}
(Integrable, \cite{Silva2008}). 
An almost complex structure $J$ on a manifold $\cM$ is referred to as 
integrable if and only if $J$ is induced by a structure of complex manifold on 
$\cM$. 
\end{Def}
\begin{Def}
(K\"ahler manifold, \cite{Kobayashi1969}).  
Let $(\cM,g,J)$ be an almost K\"ahler manifold. 
If $J$ is integrable, then $(\cM,g,J)$ is referred to as 
a K\"ahler manifold.
\end{Def}
\begin{Lemma}
\label{fact-when-K"ahler-manifold}
Let $(\cM,g,J)$ be an almost Hermite manifold, and $\nabla$ the 
Levi-Civita connection. Then it follows that \cite{Moroianu2004} 
$$
\nabla J=0\qquad\Longleftrightarrow\qquad 
\mbox{$(\cM,g,J)$ is a K\"ahler manifold.} 
$$
\end{Lemma}
Symplectic manifolds are of interest in even dimensional manifolds.  
They are defined as follows.
\begin{Def}
(Symplectic structure and symplectic manifold, \cite{Arnold}).    
Let $\cM$ be a manifold, and 
$\omega$ a two-form such that it is closed  ($\dr\omega=0$) 
and  non-degenerate. Then $\omega$ is referred to as 
a symplectic (two-) form or a symplectic structure. 
In addition $(\cM,\omega)$ is referred to as a symplectic manifold. 
\end{Def}
The dimension of symplectic manifolds is even, and 
there exist special coordinates.
\begin{Def}
(Canonical coordinates, \cite{Arnold}).     
Let $(\cM,\omega)$ be a $2n$-dimensional 
symplectic manifold. If the coordinates 
$(q,p)$ with $q=\{q^{\,1},\ldots,q^{\,n}\}$ and 
$p=\{p_{\,1},\ldots,p_{\,n}\}$ are such that
$$
\omega=\sum_{a=1}^{n}\dr p_{\,a}\wedge \dr q^{\,a},
$$
then $(q,p)$ are referred to as the canonical coordinates.
\end{Def}
Symplectic vector fields on symplectic manifolds play roles. 
They are defined as follows.
\begin{Def}
(Symplectic vector field, \cite{Silva2008}).    
Let $(\cM,\omega)$ be a symplectic manifold, and $X$ a vector field.
If $\cL_{\,X}\omega=0$, then $X$ is referred to as a symplectic vector field. 
\end{Def}
Symplectic connections on symplectic manifolds play roles. They are defined as 
follows. 
\begin{Def}
(Symplectic connection, \cite{Bieliavsky2006}).    
Let $(\cM,\omega)$ be a symplectic manifold, and $\nabla$ a connection.
If the two conditions (i)\, $\nabla\,\omega=0$ and (ii)\, $T^{\,\nabla}=0$ are 
satisfied, 
then $\nabla$ is referred to as a symplectic connection. 
\end{Def}

It is known a sufficient condition for statistical manifolds to admit 
a symplectic two-form. 
\begin{Lemma}
\label{fact-noda-when-statistical- manifold-induces-symplectic-two-form}
(Noda, \cite{Noda2011}).  
Let $(\cM,g,\nabla)$ be a statistical manifold, and 
$J$ an almost complex structure such that $(\cM,g,J)$ is an almost 
Hermite manifold, $g(JX,JY)=g(X,Y)$.  
If the following conditions:  
$$
\nabla^{\,*}_{\,X}Y
=\nabla_{\,X}Y-J(\nabla_{\,X}J)Y,\quad \mbox{and}\quad
(\nabla_{\,X}J)Y
=(\nabla_{\,Y}J)X,
$$
for all $X,Y\in T\cM$ are satisfied, 
then the two-form 
$\omega$ defined by \fr{Kahler-form} 
is a symplectic two-form, and $\nabla$  a symplectic connection. 
\end{Lemma}
\begin{Remark}
\label{fact-symplectic-statistical-manifold-Levi-Civita}
If $\nabla$ is the Levi-Civita connection and $\nabla J=0$, 
then applying this Lemma, one has that 
$\omega$ is a symplectic two-from and $\nabla$ a symplectic 
connection. 
\end{Remark}

One defines the following manifold, where it admits a symplectic structure 
and a symplectic connection.  
\begin{Def}
\label{definition-symplectic-statistical-manifold}
(Symplectic statistical manifold, \cite{Noda2011}).   
Let $(\cM,g,\nabla)$ be a statistical manifold, 
$(\cM,g,J)$ an almost K\"ahler manifold. If 
$\nabla\omega=0$ with $\omega$ defined by \fr{Kahler-form}, 
then $(\cM,g,J,\nabla)$ is referred to as 
a symplectic statistical manifold.
\end{Def}
\begin{Remark}
\label{fact-when-Killing-symplectic}
If a vector field $K$ satisfies $\cL_{\,K}g=0$ and $\cL_{\,K}J=0$, then 
it follows from Lemma\,\ref{fact-almost-Kahler-Killing} that $K$ is
a symplectic vector field. 
\end{Remark}
Similar to symplectic statistical manifold defined in 
Definition.\,\ref{definition-symplectic-statistical-manifold}, 
some other manifolds have been proposed. These include  
holomorphic statistical manifold \cite{Furuhata2009}.

The following is a relevant example of symplectic statistical manifold.
\begin{Example}
\label{example-Poincare-upper-half-model-Kahler-like}
(A generalized Poincar\'e upper half-plane model as a symplectic statistical manifold).  
Let $\rmH$ be the upper half-plane introduced in 
Section\,\ref{section-parameter-maps},  
$(x,y)$ its coordinates, and 
$g_{\,0}$ a function on $\rmH$. 
Put $g$ and $J$ to be   
$$
g=g_{\,0}(x,y)\dr x\otimes \dr x+g_{\,0}(x,y) \dr y\otimes \dr y, 
\qquad
J
=\dr y\otimes \frac{\partial}{\partial x}-\dr x\otimes \frac{\partial}{\partial y}, 
$$
so that $(\cM,g,J)$ is an almost Hermite manifold.
Let $\{\,\Gamma_{\,ab}^{\ \ c}\,\}$, $(a,b,c\in \{x,y\})$ 
be a set of connection coefficients such that 
$\nabla_{\partial/\partial \zeta^{\,a}}(\partial/\partial\, \zeta^{\,b})
=\Gamma_{\,ab}^{\ \ c}(\partial /\partial\, \zeta^{\,c})$, where $\zeta^{\,x}=x$ 
and $\zeta^{\,y}=y$. 
Choose the connection to be the Levi-Civita one. 
In this case one has  
\beqa
\Gamma_{xx}^{\ \ x}
&=&\Gamma_{x},\quad
\Gamma_{xy}^{\ \ x}
=\Gamma_{yx}^{\ \ x}
=-\Gamma_{y},\quad
\Gamma_{yy}^{\ \ x}
=-\Gamma_{x},
\non\\
\Gamma_{xx}^{\ \ y}
&=&\Gamma_{y},\quad
\Gamma_{xy}^{\ \ y}
=\Gamma_{yx}^{\ \ y}
=\Gamma_{x},\quad
\Gamma_{yy}^{\ \ y}
=-\Gamma_{y},
\non
\eeqa
where $\Gamma_{x}$ and $\Gamma_{y}$ are the following functions on $\rmH$
$$
\Gamma_{\,x}
=\frac{1}{2g_{\,0}}\frac{\partial g_{\,0}}{\partial x},\qquad\mbox{and}\qquad 
\Gamma_{\,y}
=\frac{-1}{2g_{\,0}}\frac{\partial g_{\,0}}{\partial y}.
$$
Applying Lemma\,\ref{fact-statistical-manifold-Levi-Civita}, one has 
that $\nabla^{\,*}=\nabla$. It is straightforward to verify that $\nabla J=0$.
By applying Lemma\,\ref{fact-when-K"ahler-manifold} with $\nabla J=0$, 
one has that 
$(\rmH,g,J)$ is a K\"ahler manifold. Also, 
by noticing  
Remark\,\ref{fact-symplectic-statistical-manifold-Levi-Civita}, one has  
that $(\rmH,g,J,\nabla)$ is a symplectic statistical manifold. Thus  
the two-form 
$$
\omega
=-g_{\,0}\,\dr x\wedge \dr y,
$$
constructed such that $\omega(X,Y)=g(JX,Y)$ for all $X,Y\in T\rmH$,  
is a symplectic form.
In addition, $(q,p)$ with $q=x,p=G_{\,0}(x,y)$ 
is the set of canonical coordinates, where $G_{\,0}$ is such that 
$\partial\, G_{\,0}/\partial y=g_{\,0}(x,y)$. 
To verify that all the Killing vector fields are 
symplectic vector fields, let 
$K=K^{\,x}\partial/\partial x+K^{\,y}\partial/\partial y$ be 
a Killing vector field, and 
$Z=Z^{\,x}\partial/\partial x+Z^{\,y}\partial/\partial y$ 
a vector field such that $\cL_{\,Z}J=0$. 
Then the functions $K^{\,x}$ and $K^{\,y}$ should 
satisfy 
$$
K^{\,x}\frac{\partial g_{\,0}}{\partial x}
+K^{\,y}\frac{\partial g_{\,0}}{\partial y}
+2g_{\,0}\frac{\partial K^{\,x}}{\partial x}
=0,\qquad
\frac{\partial K^{\,x}}{\partial y}+\frac{\partial K^{\,y}}{\partial x}
=0,\qquad
\frac{\partial K^{\,x}}{\partial x}-\frac{\partial K^{\,y}}{\partial y}
=0,
$$
and $Z^{\,x},Z^{\,y}$ should satisfy 
$$
\frac{\partial Z^{\,x}}{\partial y}+\frac{\partial Z^{\,y}}{\partial x}
=0,\qquad
\frac{\partial Z^{\,x}}{\partial x}-\frac{\partial Z^{\,y}}{\partial y}
=0.
$$
Thus all the Killing vector fields satisfy $\cL_{\,K}J=0$.
Combining this with Remark\,\ref{fact-when-Killing-symplectic}, one has that 
$\cL_{\,K}\omega=0$.
\end{Example}

Then one has the following main theorem in this paper 
on the phase space of the parameter maps.
\begin{Thm}
(Phase space of the parameter maps).    
Let $\rmH$ be the upper half-plane,   
$\{\cF_{\,\alpha}\}$ the parameter map family on $\rmH$,   
$g^{\,\F}$ 
the Fisher metric tensor field for the Cauchy distribution  
$C(\xi;\nu,\gamma)$ given in 
\fr{Fisher-metric-Cauchy-nu-gamma}, $\nabla$ the Levi-Civita connection 
and $J$ the almost complex structure defined in 
Example\,\ref{example-Poincare-upper-half-model-Kahler-like}.  
Then $(\rmH,g^{\,\F},J,\nabla)$ is a symplectic statistical manifold,  
$\omega=-1/(2\gamma^{\,2})\,\dr\nu\wedge\dr\gamma$  a symplectic form, and 
$(q,p)$ with $q=\nu$ and $p=1/(2\gamma)$  a set of canonical coordinates. 
The parameter maps $\{\cF_{\,\alpha}\}$ are dynamical systems on this 
symplectic statistical manifold. 
Also, it follows that $\cL_{K_{\,1}}\omega=\cL_{K_{\,2}}\omega=\cL_{K_{\,3}}\omega=0$, 
where $K_{\,1},K_{\,2}$ and $K_{\,3}$ have been defined in  
\fr{Killing-Poincare-metric}. 
\end{Thm}
\begin{Proof}
In Example\,\ref{example-Poincare-upper-half-model-Kahler-like},  
choose $x=\nu$, $y=\gamma$, and $g_{\,0}(x,y)=1/(2y^{\,2})$. Then  
one completes the proof.
\qed
\end{Proof}

With the canonical coordinates $(q,p)$, one can write 
the parameter map family $(q,p)\mapsto (q^{\,\prime},p^{\,\prime})$ as 
$$
q^{\,\prime}
=\alpha\,q\,
\frac{\left(\frac{1}{2p}\right)^2+q^{\,2}-1}{\left(\frac{1}{2p}\right)^2+q^{\,2}},
\qquad 
p^{\,\prime}
=\frac{p}{\alpha}
\,\frac{\left(\frac{1}{2p}\right)^2+q^{\,2}}
{\left(\frac{1}{2p}\right)^2+q^{\,2}+1}.
$$
In these coordinates, the non-symplectic property of the parameter maps 
is clear: 
$\dr q^{\,\prime}\wedge\dr p^{\,\prime}\neq \dr q\wedge \dr p$.

Note that when symplectic statistical manifolds are  dually flat, 
the corresponding affine coordinates can be treated as canonical coordinates 
\cite{Noda2011}.

\section{Conclusions}
In this paper,  
the parameter maps have been derived 
by reducing the Perron-Frobenius equations for the 
generalized Boole transform family without any approximation.  
For the parameter maps, it has been found that 
the statistical picture and the orbital picture are related in terms of 
complex variables. Since the dimension of the parameter space 
expressing distribution functions is even,  
it has been natural to discuss information geometry and symplectic geometry.  
Then the derived parameter map family has geometrically 
been characterized. 

There are some potential future studies that follow from this paper. 
One is to apply the present approach to other solvable chaotic systems. 
Since this study has been restricted 
to a particular family of maps, 
it is interesting to see if this approach can be extended to 
other maps, including higher dimensional maps.
To this end, one needs parameter maps for given (solvable) maps. 
However, as it can be guessed from this work, 
it may not be straightforward to obtain such parameter maps from given 
maps.
Thus, to develop a general theory, a sophisticated and systematic 
manner is demanded for obtaining parameter maps.  
After establishing such a systematic manner, one 
will investigate a parameter 
map on a statistical manifold. Also, a combination of 
this statistical manifold and symplectic or contact manifolds is expect to 
be a basis for discussions.   
If the dimension of a derived parameter map is even, 
approaches based on this work on symplectic and related 
structures are expected to give fruitful benefits, since 
known tools in symplectic and related geometries will 
be useful for characterizing parameter maps on statistical manifolds.      
These structures include conformal symplectic one \cite{Bazzoni,DO98}.
Besides, if the dimension of the parameter map is odd, then contact geometry 
will be expected to play a 
role, since contact geometry is an odd-dimensional counterpart of symplectic 
geometry \cite{Silva2008,Arnold}. 

\section*{Acknowledgments}
The authors  would like to thank 
M. Shintani, and K. Okubo (Kyoto University)
for stimulating discussions.


\end{document}